\documentclass{PoS}

\def\lesssim{\buildrel < \over {_{\sim}}}

\def\nn{\nonumber}

\title{Propagation of cosmic rays into diffuse clouds}

\ShortTitle{Propagation of cosmic rays into diffuse clouds}

\author{\speaker{Giovanni Morlino}\\
        INFN /Gran Sasso Science Institute, viale F. Crispi 7, 67100 L'Aquila, Italy \\
        APC, AstroParticule et Cosmologie, Universit\'e Paris Diderot, Paris, France \\
        INAF/Osservatorio Astrofisico di Arcetri, Largo E. Fermi, 5, 50125, Firenze, Italy \\
        E-mail: \email{giovanni.morlino@gssi.infn.it}}

\author{Stefano Gabici\\
        APC, AstroParticule et Cosmologie, Universit\'e Paris Diderot, CNRS/IN2P3, CEA/Irfu, Observatoire de Paris, Sorbonne Paris Cit\'e, 10, rue Alice Domon et L\'eonie Duquet, F-75205 Paris Cedex 13, France \\
        E-mail: \email{stefano.gabici@apc.univ-paris7.fr}}

\abstract{We study the capability of low-energy cosmic rays (CR) to penetrate into diffuse clouds when they move from the hot ionized plasma to a cool cloud embedded in that plasma. 
The spectrum of CR inside a cloud can be remarkably different from the the one present in the hot interstellar medium because when CRs pass through a dense cloud of matter, they suffer energy losses due to ionization and nuclear interactions. Hence there is a net flux of CRs towards the cloud that can excite  Alfv\'en waves. In turn, self-excited Alfv\'en waves enhances the diffusion of CRs near the edge of the cloud, forcing CRs to spend more time in this layer and increasing the amount of energy losses. The final effect is that the flux of CR entering into the cloud is strongly suppressed below an energy threshold whose value depends on ambient parameters.
For the first time we use the full kinetic theory to describe this problem, coupling CRs and Alfv\'en waves through the streaming instability, and including the damping of the waves due to ion-neutral friction and the CR energy losses due to ionization and pion production.
Differently from previous approaches, this method allow us to describe the full CR spectrum and the magnetic turbulence spectrum in the transition region between hot interstellar medium and the cloud.
For the typical size and density of diffuse clouds in the Galaxy, we find that the flux of particles entering the cloud is strongly reduced below $E\sim 10-100$ MeV, while for larger energies the CR spectrum remain unaltered.}

\FullConference{Cosmic Rays and the InterStellar Medium - CRISM 2014,\\
		24-27 June 2014\\
		Montpellier, France}

\begin{document}

\section{Introduction} \label{sec:intro}
The interaction between cosmic rays (CR) and clouds has been studied for many years because two main reasons: predicting the ionization rate induced by CRs  and understanding the gamma ray emission due to hadronic collisions occurring between CRs and the material inside the cloud. Both problems require to understand how CRs enter the clouds and how they propagate inside them. 
In this paper we focus on the penetration of CRs when they propagate from the hot phase of the interstellar medium (HISM), whose typical density and temperature are $n\sim 0.01-0.1$ cm$^{-3}$ and $T\sim 10^4-10^5$K, towards a diffuse cloud (with $n\sim 10-100$ cm$^{-3}$ and $T\sim 10^{3}$K). We concentrate only on diffuse clouds  because denser molecular clouds are usually embedded into diffuse clouds.

The first systematic studies on how CRs interact with clouds have been done in  \cite{Skill-Strong76, Ces-Voelk77}. 
In passing from the HISM toward a cold cloud, the propagation properties of CRs are affected by the presence of neutral Hydrogen (both atomic and molecular) through two different effects: {\it i}) CRs lose energy through ionization and {\it ii}) the ion-neutral friction due to Hydrogen damps the magnetic turbulence which scatters CRs. As a consequence a non linear chain of processes takes place. The authors of \cite{Skill-Strong76} first realized that because of energy losses, the cloud acts as a sink of CRs and a net flux of CRs is established towards the cloud. The resulting density gradient of CRs can excite Alfv\'enic turbulence through streaming instability. On the other hand the excited magnetic turbulence is also damped by ion-neutral friction. This damping will eventually dominate inside the cloud, but in the transition layer between the HISM and the cloud, the dominant process between amplification and damping will depend on the CR spectrum as well as on the Hydrogen density profile. As a consequence of the whole process, \cite{Skill-Strong76} found that CRs with energy $E \lesssim 300$ MeV do not penetrate into the cloud. Similar result has been obtained by \cite{Ces-Voelk77}, but with a lower energy threshold, namely $\lesssim 50$ MeV. We note that both works approximate the cloud density as a step function, hence they provide no information on the spatial behavior and predict that the magnetic amplification occurs only on the outskirt of the cloud. 
More recently \cite{EZ11} reanalyzed the same problem with a 1D hydrodynamic approach, where CRs are described as a fluid and the coupling between self-excited waves and CRs is treated self-consistently. In this work the authors conclude that the CR pressure in the cloud drops only by $\sim 8\%$ and ascribe such decrease to the fact that CRs decouple from the plasma because of the wave damping. Nevertheless the hydrodynamic approach can only handle with integrated quantities and do not allow one to have detailed information on the CR spectrum. 
In order to overcome this drawback, here we examine the propagation problem using a kinematical description for both CRs and magnetic Alfv\'enic turbulence. The diffusive transport equation of cosmic rays is coupled to the magnetic turbulence which determines the diffusion properties of the particles and, in turn, the Alfv\'en waves are amplified in presence of CR density gradient. Our approach can be thought as an extension of all previous works. With respect to \cite{Skill-Strong76} and \cite{Ces-Voelk77} we can describe completely the space dependence of the CRs and turbulence, and with respect to \cite{EZ11} we also provide information on the energy spectra. 
In \S~\ref{sec:model} we introduce the mathematical description of the model while in \S~\ref{sec:results} we discuss our results.

\section{The mathematical model}
 \label{sec:model}
\begin{figure}
\begin{center}
{\includegraphics[width=0.4\textwidth]{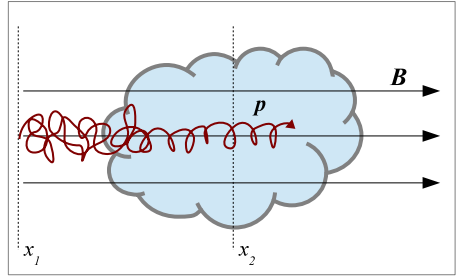}}
\end{center}
\caption{Schematic picture of CR diffusing through a diffuse clouds along the direction of the large-scale magnetic field $\bf B$. Deep inside the cloud, where the magnetic turbulence has been totally damped, we do expect a spiral motion along the field lines.}
\label{fig:sketch}
\end{figure}

The propagation of CRs through the HISM is thought to be diffusive, a fact which is strongly supported by many observations. Particles are scattered by the magnetic turbulence present in the Galaxy, mainly Alfv\'en waves. In order to describe the process of CR propagation from the HISM towards a diffuse cloud we use the same diffusive approximation. The validity of such approximation inside the cloud is questionable, because the magnetic turbulence should be damped very efficiently by the ion-neutral friction. Nevertheless here we are mainly interested in the description of the propagation in the transition region between the hot interstellar medium and the cloud. In this region the density of neutral hydrogen is not highly enough to damp completely the Alfv\'enic turbulence and the diffusive approximation can still hold. 
We will use a 1D model, which is justified by the fact that CRs diffuse mainly along the direction of the magnetic field, and the typical coherence length of the magnetic field in the Galaxy is 50--100 pc, while the typical dimension of a diffuse cloud is only $\sim 10$ pc. The geometry of the problem is sketched in Fig.~\ref{fig:sketch}. The strength of the large scale component of the magnetic field is assumed to be constant also inside the cloud, a fact which is supported by observations \cite{Crutcher10} showing that for low density ISM, the magnetic field strength is independent of the ISM density up to a density $n<300$ cm$^{-3}$. 

In the following subsections we will illustrate how we describe each single piece of the problem: the transport of CRs, the transport of Alfv\'en turbulence, the ion-neutral damping and the resonant amplification due to CR streaming.  

\subsection{Transport equation for Cosmic Rays} \label{sec:CRs}
The one dimensional steady state equation for the transport of CRs in presence of diffusion, advection and energy losses, written in the rest frame of the plasma reads 
\begin{equation}
  \frac{\partial}{\partial x} \left[ D \frac{\partial f}{\partial x} \right] - v_A \frac{\partial f}{\partial x} 
   + \frac{p}{3} \frac{\partial v_A}{\partial x} \frac{\partial f}{\partial p} 
   - \frac{1}{p^2} \frac{\partial}{\partial p} \left[ \dot p p^2 f \right] = 0 \,,
\label{eq:fCR}
\end{equation}
where $f(x,p)$ is the particle distribution function, $D(x,p)$ is the diffusion coefficient, $\dot p(x,p)$ is the momentum losses and $v_A(x)= B(x)^2/\sqrt{4\pi \rho_i}$ is the Alfv\'en speed in the regular magnetic field $B(x)$.  Notice that in this expression we include only the ions mass density, $\rho_i$, because at the wavelengths relevant in this work, the wave frequency is smaller than the charge-exchange frequency between ions and neutrals, hence, while ions oscillate with waves, neutrals have no time to couple with them (see e.g.\cite{ZS82} for more details).

Concerning the diffusion process, we consider only the effect produced by Alfv\`enic turbulence which gives the following diffusion coefficient \cite{Bell78a}
\begin{equation} 
  D(x,p)= \frac{4}{3\pi} \frac{r_L(p)}{\mathcal{P}_w(x, \bar k(p))/P_B} \,,
\label{eq:D}
\end{equation}
where $r_L(p)$ is the Larmor radius of particles in the large-scale magnetic field $\mathbf B$ and $P_B= B^2/(8\pi)$ is the magnetic pressure of the regular field $\mathbf{B}$. $\mathcal{P}_w(x, k)$ is the magnetic pressure in turbulent mode with wavenumber $k$ and will be derived in the \S\ref{sec:waves}. Notice that Eq.(\ref{eq:D}) is strictly valid only in the linear regime, namely when $\mathcal{P}_w(x, k)/P_B \ll 1$. We checked {\it a posteriori} that this condition is always verified in our results.

In the region deep inside the cloud where the magnetic turbulence is totally damped, we set $D=L v(p)/3$ where $L$ is the cloud dimension and $v(p)$ the particle speed. Using this trick we can still use the diffusive approach which gives the correct travel time for the cloud crossing.

We limit our calculation to CRs with energies between 100 keV and 10 GeV. In this energy region the most important channels of energy losses are due to ionization of neutral hydrogen atoms and, for $E>1$ GeV, also to the pion production. We took the energy losses from \cite{Padovani09} (see their Fig.(7)) and we use an analytic expression which well fit the data in the energy range [100 keV, 10 GeV ]:
\begin{equation}
  \dot{p} /(m_p c) =  \left( n_H/{\rm cm^{-3}} \right)  \left\{ k_1 \left( p/p_1 \right)^{\alpha_1} 
  	\left[ 1 + \left( p/p_1 \right)^{\gamma(\alpha_1-\alpha_2)} \right]^{-1/\gamma} 
	+ k_2 \left( p/p_2 \right)^{\alpha_3} \right\} {\rm s^{-1}}\,.
\label{eq:losses}
\end{equation}
The fit procedure gives the following value for the parameters: $k_1=7.3\cdot 10^{-13}$, $k_2=9.5 \cdot 10^{-17}$, $p_1=0.011\, m_p c$, $p_2=0.503\, m_p c$, $\alpha_1=-1.58$, $\alpha_2=1.197$, $\alpha_3=1.1598$ and $\gamma= 1.2$.

Let us now discuss the boundary conditions. Outside the cloud we impose that the CR distribution reduces to the Galactic one, hence the boundary condition at $x=x_1$ is $f(x_1,p)=f_0(p)$. We also impose a second boundary condition in the centre of the cloud, where for symmetry we have $\partial_x f(x,p)|_{x=x_2} = 0$. This boundary condition is an important difference with respect to previous works, which impose the second boundary condition on the CR gradient far from the cloud. While \cite{Skill-Strong76} and \cite{Ces-Voelk77} set $\partial_x f(x,p)|_{x=x_2} = 0$, in \cite{EZ11} the authors set this same gradient equal to a fixed number. In our case, on the other hand, the value of $\partial_x f(p,x)|_{x=x_2}$ is an outcome of the calculation.
We stress that the symmetry condition is the most meaningful boundary condition and it is valid as long as the diffusion approximation holds. On the other hand, when the diffusion approximation loses validity, the distribution function is not in general isotropic but will depend on the pitch angle $\alpha$. In this case the symmetry condition becomes $f(x_2,p,\alpha) = f(x_2,p,\pi-\alpha)$.

Eq.~(\ref{eq:fCR}) cannot be solved in an explicit way because it is nonlinear, due to the dependence of the diffusion coefficient on the magnetic turbulence which is, in turn, determined by the gradient of the CR density. In order to get the solution we  implement an iterative technique similar to those used to solve the nonlinear problem of shock acceleration (see e.g.\cite{AmatoBlasi05, Morlino13}). 
The first step is to write an implicit solution for Eq.~(\ref{eq:fCR}). We define the function $g(x,p)= D \,{\partial_x f}$ and we note that the equation for $g(x,p)$ can be formally reduced to a first order differential equation, i.e.
\begin{equation} \label{eq:g}
  \partial_x g  - \left( v_A/D \right) \, g + Q = 0 \,,
\end{equation}
with the appropriate boundary condition $g(x_2,p)=0$. The nonlinearity of the problem has been hided in the function $Q(x,p)$ which plays the role of a source/sink term, and reads
\begin{equation} \label{eq:Q}
  Q(x,p) = (p/3) \partial_x v_A \, \partial_p f - p^{-2} \partial_p \left[ \dot p p^2 f \right] \,.
\end{equation}
The implicit solution for $g$ is
\begin{equation} 
  g(x,p) = \int_{x}^{x_2} Q(x',p) \exp\left[-\int_{x}^{x'} \frac{v_A}{D(y,p)} dy \right] dx' \,.
\label{eq:g_sol}
\end{equation}
Eq.(\ref{eq:g_sol}) allows us to write down the solution for $f(x,p)$ in the following implicit form:
\begin{equation} 
  f(x,p) = f_0(p) + \int_{x_1}^{x} \frac{dx'}{D(x',p)} \int_{x'}^{x_2} Q(x'',p) 
    \exp\left[-\int_{x'}^{x''} \frac{v_A}{D(y,p)} dy \right] dx'' \,.
\label{eq:sol_fCR}
\end{equation}
This equation can be solved iteratively, once the functions $D$, $v_A$ and $\dot p$ are fixed. Nevertheless we note that a special care should be taken in choosing the solving algorithm, because the convergence of Eq.~(\ref{eq:sol_fCR}) is not guaranteed a priori.

\subsection{Transport equation for Alfv\`en waves} \label{sec:waves}
Whenever charged particles stream faster than the Alfv\'en speed, they generate Alf\'en waves with wavelengths close to their Larmor radius. In the steady state approximation and for one dimensional system, the equation for the transport of magnetic turbulence reads as (see e.g. \cite{McKenzieVoelk82})
\begin{equation}
  \partial_x \mathcal{F}_w(x,k) =  \sigma(x,k) - \Gamma(x,k) \, \mathcal{P}_w(x,k) \,,
\label{eq:Fw}
\end{equation}
where  $\mathcal{F}_w(x,k)$ and $\mathcal{P}_w(x,k)$  are, respectively, the energy flux and pressure per unit logarithmic bandwidth of waves with wavenumber $k$. Notice that we are considering only waves moving towards the cloud, hence only with one polarity. $\sigma(x,k)$ is the growth rate of energy in magnetic turbulence and $\Gamma(x,k)$ is the rate at which the turbulence is damped. In principle, in Eq.~(\ref{eq:Fw}) we should include also the wave-wave coupling term, which produce a diffusion in the $k$-space, but here we neglect such a term because the damping rate due to neutral friction and the growth rate due to streaming instability are both faster than the wave-wave damping.
For Alfv\'en waves the following relations hold:
\begin{eqnarray}
  \int dk \, k^{-1} \mathcal{P}_w(x,k)  &=&         (\delta B)^2/(8\pi)  \equiv P_w \,, \nn \\
  \int dk \, k^{-1} \mathcal{F}_w(x,k)  &=&  v_A (\delta B)^2/(4\pi)  \equiv F_w \,.
\label{eq:PwFw}  
\end{eqnarray}
such that the relation between energy flux and pressure is $\mathcal{F}_w= 2 v_A\, \mathcal{P}_w$ \cite{Caprioli08}. Using this relation Eq.(\ref{eq:Fw}) reduces to the following expression
\begin{equation} \label{eq:Pw}
  \partial_x \mathcal{P}_w(x,k) = - \left[ \partial_x \ln(v_A) + \Gamma/(2 v_A) \right]  \mathcal{P}_w + \sigma/(2 v_A)
\end{equation}
which can be easily solved once the boundary condition far from the cloud is specified. We set $\mathcal{P}_w(x_1,k) = \mathcal{P}_{w,0}(k)$, where the choice of $\mathcal{P}_{w,0}(k)$ presents some subtleties which will be discussed in \S~\ref{sec:boundary}. Using this boundary condition, the solution of Eq.(\ref{eq:Pw}) reads
\begin{equation} 
  \mathcal{P}_w(x,k) = \mathcal{P}_{w,0}(k) + \frac{1}{2 v_A} \int_{x_1}^x \left[ \sigma - \mathcal{P}_{w,0}(k) 
     \left( \Gamma + 2   \frac{\partial v_A}{\partial x'} \right) \right] 
     \exp \left\{ - \int_{x'}^x \frac{\Gamma}{2 v_A} dy \right\} dx'
\label{eq:Pw_sol}
\end{equation}
To understand the physical meaning of each term, it is useful to look at the solution (\ref{eq:Pw_sol}) in the case where $v_A$, $\Gamma$ and $\sigma$ are all constant in space. In this limit the solution becomes
\begin{equation} 
  \mathcal{P}_w(x,k) = \mathcal{P}_{w,0}(k) e^{-\frac{\Gamma}{2 v_A}(x-x_1)} + \sigma/\Gamma \, 
    \left( 1- e^{-\frac{\Gamma}{2 v_A}(x-x_1)}\right) \,,
\label{eq:Pw_const}
\end{equation}
where we recognize a local term, $\sigma/\Gamma$, plus a propagation term $\propto e^{-\Gamma x/v_A}$. In the further limit where the Alfv\`en speed is negligible with respect to the damping speed, $\Gamma x$, the propagation of waves drops to zero, and the power spectrum is determined only by the local ratio between the growth and the damping rates, i.e. $\mathcal{P}_w(x,k) = \sigma(x,k)/\Gamma(x,k)$.

We still need to explicit the amplification and the damping rates. The only source of magnetic amplification is the resonant scattering between the accelerated particles and the magnetic turbulence, which gives the following growth-rate \cite{Skilling75}
\begin{equation} \label{eq:sigma}
 \sigma(x,k) = (4\pi/3) v_A(x) \left[ p^4 v(p) \partial_x f \right]_{p=\bar p(k)} \,,
\end{equation}
where the term in the square brackets is calculated for $\bar p(k)= eB/(k m_p c)$, which is the well know resonant condition between particles and Alfv\'en waves.
The main source of damping is, instead, the ion-neutral friction. In the present work we deal only with diffuse clouds, whose typical temperature and density are 10--100 K and 10--100 cm$^{-3}$, respectively, and are dominated by atomic hydrogen. Hence we only need to account  for the elastic scattering between protons and atomic hydrogen. Moreover, for the particle energies we are interested in ($E< 10$ GeV), the ion-neutral damping does not depend on the wave number $k$, but it is only a function of the position. The resulting damping rate is \cite{Kuls-Cesa71,Drury96}:
\begin{equation} \label{eq:damping}
 \Gamma = 4.2 \times 10^{-9} \left(T/10^4 \rm K \right)^{0.4} \left( n_H/{\rm cm^{-3}} \right) \, {\rm s^{-1}}\,,
\end{equation}
where $T$ is the temperature o the plasma and $n_H$ is the atomic Hydrogen density.

\subsection{Boundary conditions far from the cloud} \label{sec:boundary}

In order to solve simultaneously Eq.~(\ref{eq:sol_fCR}) and Eq.~(\ref{eq:Pw_sol}) one has to chose self consistently the boundary conditions $f_0(p)$ and $\mathcal{P}_{w,0}(k)$ far from the cloud, in such a way to match the conditions in the Galaxy. The Local Galactic CR spectrum is well know only for energies above few GeV, where $f(p)\propto p^{-4.7}$. For lower energies the spectrum is strongly affected by the solar modulation, which prevent us to know the actual Galactic spectrum. There are both theoretical and experimental reasons to believe that for energies below few hundreds MeV the slope of CR spectrum becomes harder. Without any loss of generality we chose here a simple broken power law:
\begin{equation} \label{eq:f0}
  f_0(p) = K_f \left(p/p_0\right)^{-s_1} \left[ 1 + \left(p/p_0\right)^{s_1-s_2} \right]^{-1}  \,,
\end{equation}
where $s_1= 4.7$ is the slope for $p>p_0\equiv 0.2$ GeV, while $s_2$ is the slope below $p_0$ and will be taken as a free parameter. $K_f$ is the normalization constant chosen in order to have a CR pressure equal to 1 eV cm$^{-3}$.

Here it is worth to underline a subtlety of the one dimensional model we are using. From a pure mathematical point of view, the value of $\mathcal{P}_{w,0}(k)$ should be strictly zero, because in the absence of any CR gradient, no turbulence is excited. On the other hand in the Galaxy this situation is never realized, because in the absence of diffusion the CR distribution would develop a strong anisotropy (due to a pure ballistic motion) which will turn on the self-generation of turbulence. Hence a physical justified boundary condition for $\mathcal{P}_{w,0}(k)$ is the actual Galactic turbulence, which determines the Galactic CR spectrum far from the cloud. The magnetic turbulence spectrum in the Galaxy is not well known, but in literature the most widely used one is the Kolmogorov power spectrum, which reads 
\begin{equation} \label{eq:Pw0}
  \mathcal{P}_{w,0}(k) = 2/3 \, \eta_w P_{B,0} \left( k L_{\rm tur} \right)^{-2/3} \,.
\end{equation}
The normalization factor $\eta_w$ gives the power in the magnetic turbulence with respect to the power in the regular field, and it is chosen in order to fix the diffusion coefficient at 1 GeV equal to $10^{28} \rm cm^2 s^{-1}$. This condition gives $\eta_w=0.6$.  $L_{\rm tur}$ is the maximum scale at which the turbulence is injected into the Galaxy, which is typically assumed to be between 50 and 100 pc. In this work we use the value 50 pc. Using Eq.~(\ref{eq:D}) we see that the Kolmogorov turbulence implies a diffusion coefficient $D(p) \propto p^{1/3}$.

\subsection{Spatial profile of the cloud} \label{sec:cloud}
To study the penetration of CRs into a diffuse cloud we need to specify how the density profiles of both neutral hydrogen and ionized protons change in passing from the HISM to the cloud interior. We adopt a profile similar to what has been used in \cite{EZ11}. The total density profile of the plasma is described by a tanh profile:
\begin{equation} 
    \rho(x) = \rho_1 + \left(\rho_1-\rho_2 \right) \, \left( 1+ \tanh \left[ (x-x_c)/\Delta x_c \right] \right) /2 \,,
\label{eq:rho}  
\end{equation}
where $x_c$ is the position of the edge of the cloud, and $\Delta x_c$ is the thickness of the transition. We chose $x_c=0.5$ pc and $\Delta x_c$ between 0.7 and 0.05 pc. $\rho_1 \equiv \rho(x_1)$ is the density far outside the cloud, while $\rho_2 \equiv \rho(x_2)$ is the density at the centre of the cloud. We adopt the typical values are $\rho_1= 0.01\, m_p$ cm$^{-3}$ and $\rho_2= 100\, m_p$ cm$^{-3}$. 
Always following \cite{EZ11}, we describe the ionization fraction using a similar profile:
\begin{equation} \label{eq:f_ion}
  \xi_{ion}(x) = \xi_1 + \left(\xi_1-\xi_2\right) \left( 1+ \tanh\left[ -(x-x_c)/\Delta x_c \right] \right) /2 \,.
\end{equation} 
The ionization fraction is set to drop from a maximum ($\xi_2 = 1$) far outside the cloud, to a minimum ($\xi_1 = 10^{-3}$) inside the cloud. Such values are consistent with observations, but it is worth notice that, to treat the problem in a self consistent way, we should calculate the ionization produced by the CR spectrum predicted by Eq.(\ref{eq:sol_fCR}), rather than using an {\it a priori} profile like Eq.(\ref{eq:f_ion}). We will address this issue in a future work, but we stress that such a choice does not affect our main conclusions.

Finally we fix the temperature of the hot interstellar phase to be $T=3\cdot 10^6$ K, and we determine the corresponding temperature profile, $T(x)$ imposing everywhere the condition of pressure equilibrium. 

\section{Results and discussion}
 \label{sec:results}
\begin{figure}
\begin{center}
{\includegraphics[width=0.9\linewidth]{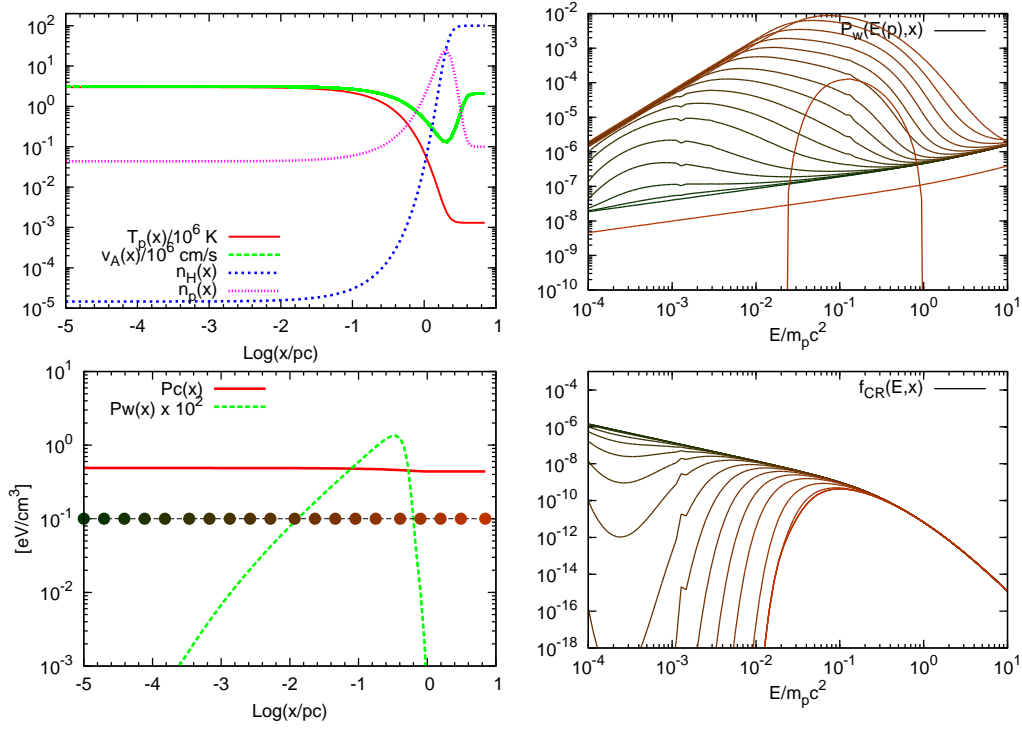}}
  \caption{{\it Top-left panel}: spatial profile of ions and Hydrogen number density, plasma temperature and Alfv\'en speed. {\it Bottom-left panel}: Total CR energy density and magnetic wave energy density ($\times 100$). The filled circles indicate the position where we take the energy spectrum of both waves and CRs which are shown in the top and bottom right panels, respectively (using the same color code). The parameters' values chosen for this case are written in the first row of Table~1.} 
 \label{fig:result1}
\end{center}
\end{figure}

In Fig.~\ref{fig:result1} we summarize the results of our calculations for a specific choice of the parameters' values, which are summarized in the first row of Table~\ref{tab:1}. The top left plot show the initial configuration of the plasma along the spatial coordinate $x$: different lines show the density profile of both neutral Hydrogen and ions, the plasma temperature, $T$, and the value of Alfv\'en speed $v_A$. Notice that the cloud size is assumed to be 10 pc.
The bottom left panel shows the results for the total CR pressure, $P_{c}(x)$, and the total energy density in magnetic waves, $P_w(x)$ (multiplied by 100). Notice that $P_w(x)$ is the integral of $\mathcal{P}_{w}(k)$ only between $k_{\min}$ and $k_{\max}$ which are the resonant wave-vectors corresponding to 10 GeV and 100 keV, respectively. We see that $P_w(x)$ start to increase far from the cloud, reaching its maximum at the location where the $n_H \approx 10^{-3}$ cm$^{-3}$ and has a value of  $\sim 10^4 \times P_{w,0}$, confirming that the amplification of the turbulence can be very effective. In spite of this fact we stress that the amplification is still in the linear regime, meaning that $\delta B/B<1$. After the peak the density of magnetic turbulence drops rapidly because the ion-neutral friction and, simultaneously, the energy density of CRs decreases of $\sim 10\%$.
The right panels show the magnetic energy spectrum, $\mathcal{P}_{w}(k)$ (top panel) and the CR spectrum, $f(p)$ (bottom panel) at different location along $x$, according to the color code. The corresponding positions are marked with filled circles of the same color in the bottom left panel: hence the dark green lines correspond to spectra far away from the cloud, while the red lines represent the spectrum inside the cloud.
Notice that $\mathcal{P}_{w}(k)$ is plotted versus the resonant energy $E_r$, such that waves with wave-vector $k$ resonate with particles with energy $E_r$, i.e. $r_L(E_r)=1/k$. In this way we can easily compare the energy density of waves with wave-vector $k$ with the energy density of particles which resonate with the same $k$. From the top-right plot we see that not all waves are excited at the same rate: the largest amplification occurs for those waves resonating with particles which lose energy faster. 
The bottom-right panel shows that the CR spectrum is depleted at low energies as we approach the cloud interior. Inside the cloud the spectrum has a peak at $E_{\rm peak}\approx 90$ MeV and remains unchanged for energies above $E_{\rm peak}$, while it is strongly damped below such value. $E_{\rm peak}$ is not a constant but depends mainly on two factors, namely the value of the density inside the cloud, $n_2$, and the thickness of the cloud edge, $\Delta x_c$. We also find a slight dependence on the CR spectral index at low energy, $s_2$. In Table~\ref{tab:1} we summarize the results for $E_{\rm peak}$ for few cases where we change the value of $s_2$, $\Delta x_c$ and $n_2$. In all cases we found $E_{\rm peak}$ between 10 and 100 MeV, confirming the "exclusion" effect first found in \cite{Skill-Strong76}, even if for the parameters used here the exclusion energy is slightly smaller.

\begin{table}
\caption{\label{tab:1} Value of the peak energy, $E_{\rm peak}$, in the CR spectrum inside the cloud for different values of parameters. The first row shows to values used to produce Fig.~1.}
\begin{center}
\begin{tabular}{ccccc | c}
\hline
 $ s_1$ & $s_2$ & $\Delta x_c$/pc &  $n_1$/cm$^{-3}$  &  $n_2$/cm$^{-3}$  & $E_{\rm peak}$/MeV \\
\hline
  4.7  &  2  &  0.5  &  0.01  &  100  &  90   \\  
  4.7  &  2  &  0.7  &  0.01  &  100  &  60   \\
  4.7  &  2  &  0.1  &  0.01  &  100  &  110  \\  
  4.7  &  1  &  0.5  &  0.01  &  100  &  70  \\  
  4.7  &  3  &  0.5  &  0.01  &  100  &  90  \\
  4.7  &  2  &  0.5  &  0.01  &      1  &  20  \\      
\hline
\end{tabular}
\end{center}
\end{table}

%
%

\end{document}